\documentclass[amsmath, amsfonts, superscriptaddress, twocolumn, prl]{revtex4}
\usepackage{graphicx}
\usepackage{epsfig}
\usepackage{bm}
\usepackage{dcolumn}
\usepackage{amsmath}
\usepackage{amssymb}
\usepackage{color}

\newcommand{\beg}{\begin{equation}}
\newcommand{\en}{\end{equation}}
\newcommand{\bp}{\mathbf p}

\newcommand{\bk}{\mathbf k}
\newcommand{\br}{\mathbf r}

\newcommand{\bn}{\mathbf n}

\newcommand \bel  {\begin{align}}
\newcommand \enl  {\end{align}}

\newcommand{\up}{\uparrow}
\newcommand{\dn}{\downarrow}
\newcommand{\dg}{^\dagger}

\begin{document}

\title{Inhomogeneous magnetic multiband superconductors}

\author{Maxim Dzero}
\affiliation{Department of Physics, Kent State University, Kent, Ohio 44242, USA}

\author{Alex Levchenko}
\affiliation{Department of Physics, University of Wisconsin-Madison, Madison, Wisconsin 53706, USA}

\date{\today}

\begin{abstract}
We consider a problem of superconductivity coexistence with the spin-density-wave order in disordered multiband metals. It is assumed that random variations of the disorder potential on short length scales render the interactions between electrons to become spatially correlated. 
As a consequence, both superconducting and magnetic order parameters become spatially inhomogeneous and are described by the universal phenomenological quantities, whereas all the microscopic details are encoded in the correlation function of the coupling strength fluctuations. We consider a minimal model with two nested two-dimensional Fermi surfaces and disorder potentials which include both intra- and inter-band scattering. The model is analyzed using the quasiclassical approach to show that short-scale pairing-potential disorder leads to a broadening of the coexistence region.
\end{abstract}

\maketitle

\paragraph{Introduction.}  
It is a well-known fact that generally disorder is detrimental to superconductivity. Although sufficiently small amount of potential scatterers in superconductors with isotropic pairing wave-function does not suppress the critical transition temperature and energy gap, the key result known as Anderson theorem \cite{Anderson}, Larkin and Ovchinnikov have shown in their seminal paper \cite{LO-model} that even when time-reversal symmetry is preserved the coherence peak in the density of states can be smeared by the disorder induced inhomogeneities. Although this result seems counter intuitive at first sight, one can understand it by observing that at the mean-field level their model naturally contains an effective depairing parameter. As a result, changes in the pair-potential field as well as single-particle correlation functions due to inhomogeneities are of the same form as those found earlier by Abrikosov and Gor'kov for the case of superconductors contaminated with magnetic impurities \cite{AG1961}. Furthermore, the hard gap in the spectrum gets also smeared due to optimal fluctuations of the order parameter, thus leading to the Lifshitz-type tail \cite{Lifshitz} in the subgap region. For refinements and extensions of the original ideas to $s$- and $d$-wave superconductors see Refs. \cite{Flatte,Klemm,Andersen,Skvortsov,Houzet} as well as extensive review \cite{Balatsky} and references herein.

Iron-based superconductors serve as a prime example \cite{Chubukov2012,Matsuda2014} of complex materials in which disorder seems to play a highly nontrivial role. These materials belong to a subclass of composite superconductors in which superconductivity with an isotropic $s^{\pm}$-order parameter may develop on multiple bands and it usually competes with magnetic order. There is an extensive literature on the effect of impurities on the pairing state in pnictides, see e.g. \cite{KontaniPRL2009,EfremovPRB2011,YamakawaPRB2013,MishraPRB2013,StanevPRB2014,HoyerPRB2015,BrydonArXiv2021}. Of specific interest to the present work, it is in the context of the physics of these materials that it was shown \cite{VC-PRB2011,Fernandes} that disorder may actually boost superconductivity either by changing the corresponding scattering rates or, as in the case of stoichiometric substitutions, by varying the relative anisotropy of the Fermi pockets \cite{VVC2010,Schmalian}. 

\begin{figure}[t!]
  \centering
 \includegraphics[width=3.15in]{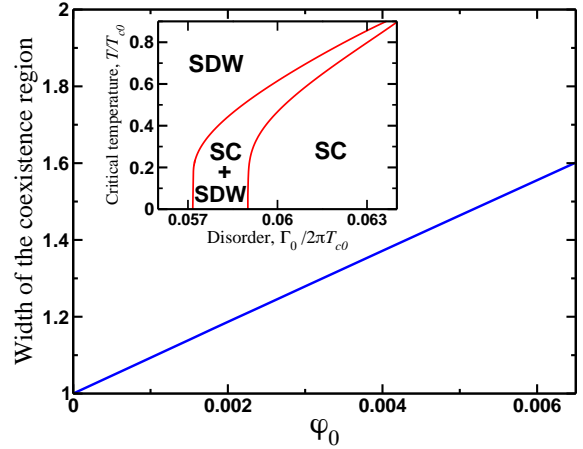}
  \caption{Main panel: schematic plot showing how the width of the coexistence region between the superconducting (SC) and spin-density-wave (SDW) orders varies as a function of the dimensionless parameter $\varphi_0$ describing the effects of spatial inhomogeneities in a system, Eqs. (\ref{g1Corr},\ref{phir}). The width is given in the units of the $\delta\Gamma_0$ which determines the one of the coexistence region in the spatially homogeneous system \cite{DzeroLondon2015}. 
  Inset: phase diagram showing a region of coexistence between SC-SDW orders as found from the mean-field analysis of the Hamiltonian (\ref{Eq1}). The boundary line to the left of the coexistence region represents the superconducting critical temperature $T_c(M)$ at finite values of magnetization while the line to the right of the coexistence region represents the critical temperature $T_s(\Delta)$ of the SDW transition at finite pairing amplitude. The temperatures are given in the units of the superconducting critical temperature in a clean system. }\label{Fig1}
\end{figure}

Behind the physical interpretation of this effect is an idea that disorder must suppress superconductivity (SC) slower than it suppresses magnetic, in that case spin-density-wave (SDW), order. Indeed, in these materials due to the $s$-wave symmetry of the pairing amplitude, Anderson theorem still partially applies in a sense that only inter-band disorder affects SC state but SDW is affected by intra-band scattering as well. This means that in the temperature-doping $(T$-$x)$ phase diagram, a narrow region of concentrations of impurity atoms must be present in which superconductivity would actually be in coexistence with SDW order. In passing we note that SC-SDW coexistence in iron-based superconductors actually leads to a number of fascinating physical effects, such as anomalous temperature and doping dependence of the heat capacity \cite{Hardy,Carrington} and London penetration depth \cite{Hashimoto,Auslaender,Joshi} near the point where the SDW vanishes and quantum critical fluctuations play a dominant role in determining their thermodynamic and transport response functions at low temperatures \cite{Levchenko2013,Chowdhury2013,Fernandes2020,Khodas2020}.

Almost proverbial antagonism between spin-singlet superconductivity and magnetism on one hand, and the possibility of their coexistence due to different disorder scattering rates on the other hand, brings up the question of whether allowing for spatial inhomogeneities of the order parameters, for example, would produce either the broadening of the coexistence region or, on the contrary, the narrowing it down. In this work we address precisely this question and show that at least within the limits of the two-band model \cite{VC-PRB2011,DzeroLondon2015} that we will adopt in what follows, the spatial inhomogeneities lead, in fact, to the broadening of the coexistence region and an enhancement of SC critical temperature. Our main result is presented in Fig. \ref{Fig1}.

\paragraph{Model.}  The Hamiltonian for the model we study below is 
\beg\label{Eq1}
\hat{\mathcal{H}}=\int{\Psi}\dg(\br)\left(\hat{H}_0+\hat{H}_{\textrm{mf}}+\hat{H}_{\textrm{dis}}\right)\Psi(\br){d^2\br}
\en
Here we use the eight-component spinor in the Balian-Werthammer representation \cite{BW-PRL63}, namely
${\Psi}_\bp\dg=\left(\hat{c}_{\bp\up}\dg, ~\hat{c}_{\bp\dn}\dg,~-\hat{c}_{-\bp\dn}, ~\hat{c}_{-\bp\up},
~\hat{f}_{\bp\up}\dg, ~\hat{f}_{\bp\dn}\dg, ~-\hat{f}_{-\bp\dn},~\hat{f}_{-\bp\up}\right),$
which contains spin-$1/2$ $c$- and $f$-fermionic fields with momentum $\bp$ and describe two (one electron- and one hole-like) bands respectively \cite{MinusSign}. $\hat{H}_0$ describes the single-particle states,  $\hat{H}_{\textrm{mf}}$ is the interaction part taken in the mean-field approximation
\beg\label{hp}
\begin{split}
\hat{H}_0&=-\xi_{\vec{\nabla}}\hat{\tau}_3\hat{\rho}_3\hat{\sigma}_0, 
~\hat{H}_{\textrm{mf}}=-\Delta{\hat{\tau}_3\hat{\rho}_1\hat{\sigma}_0}+{\bm M}{\hat{\tau}_1\hat{\rho}_0\hat{\bm{\sigma}}}.
\end{split}
\en
In the expression above $\hat{\tau}_i$, $\hat{\rho}_i$, $\hat{\sigma}_i$ are Pauli matrices operational in band/Gor'kov-Nambu/spin sub-spaces, $\xi_{\vec{\nabla}}=-{\vec \nabla}^2/(2m)-\mu$ is the single-particle dispersion, $\mu$ is a chemical potential, $\Delta$ is the superconducting order parameter and $\bm{M}$ is the magnetization which we shall take to be along the $z$-axis, $\bm{M}=M\bm {e}_z$. 
Lastly, the Hamiltonian density, which introduces disorder scattering by randomly distributed impurities in locations ${\mathbf R}_i$, is
\beg\label{Dis}
\hat{H}_{\textrm{dis}}=
\sum\limits_{i}\left[u_0(\hat{\tau}_0\hat{\rho}_3\hat{\sigma}_0)+u_\pi (\hat{\tau}_1\hat{\rho}_3\hat{\sigma}_0)\right]\delta(\br-{\mathbf R}_i).
\en
The scattering potential $u_0$ accounts for disorder scattering within each band, while the second term $u_\pi$ leads to the inter-band transitions. 

\paragraph*{Quasiclassical theory.} The ground state of the Hamiltonian described by Eq. (\ref{Eq1}) can be studied using the relatively simple system of Eilenberger equations \cite{Eilenberger}, that for the model under consideration can be cast into a single equation for the matrix function $\hat{\cal G}(\omega_n,\bn,\br)$ \cite{Kirmani}:
\beg\label{Eq2Eilenr}
\begin{split}
&\left[i\omega_n\hat{\tau}_3\hat{\rho}_3\hat{\sigma}_0;\hat{\cal G}\right]-\left[\hat{H}_{\textrm{mf}}\hat{\tau}_3\hat{\rho}_3\hat{\sigma}_0;\hat{\cal G}\right]
-\left[\hat{\Sigma}_{\omega}\hat{\tau}_3\hat{\rho}_3\hat{\sigma}_0;\hat{\cal G}\right]\\&=iv_F\left({\mathbf n}\cdot{\mbox{\boldmath $\nabla$}}\hat{\cal G}\right),
\end{split}
\en
where $\omega_n$ is the fermionic Matsubara frequency
and $[\hat{A};\hat{B}]$ represents a commutator of two matrices in each
term, respectively. The self-energy part calculated to the leading accuracy within Born approximation reads 
\beg\label{Sigma4G}
\begin{split}
\hat{\Sigma}_\omega=&-i\Gamma_{0}\hat{\tau}_3\hat{\rho}_0\hat{\sigma}_0
\hat{\cal G}\hat{\tau}_0\hat{\rho}_3\hat{\sigma}_0+\Gamma_{\pi}\hat{\tau}_2\hat{\rho}_3\hat{\sigma}_0
\hat{\cal G}\hat{\tau}_1\hat{\rho}_3\hat{\sigma}_0
\end{split}
\en
where $\Gamma_{0,\pi}=\pi n_{\text{imp}}\nu_Fu_{0,\pi}^2$ are corresponding disorder intra/inter-band scattering rates with $n_{\text{imp}}$ being the impurity concentration. Matrix function $\hat{\cal G}$ satisfies the normalization condition 
$\hat{\cal G}^2=\hat{\tau}_0\hat{\rho}_0\hat{\sigma}_0$. Equation (\ref{Eq2Eilenr}) is supplemented by the self-consistency conditions for the order parameters
\beg\label{GenSelf}
\begin{split}
&\frac{iM}{g_{\textrm{m}}}=\frac{\pi T}{8}\sum\limits_{\omega_n>0}^\Lambda \textrm{Tr}\left[(\hat{\tau}_1+i\hat{\tau}_2)(\hat{\rho}_0+\hat{\rho}_3)\hat{\sigma}_3\hat{\cal G}\right], \\
&\frac{i\Delta}{g_{\textrm{s}}}=-\frac{\pi T}{8}\sum\limits_{\omega_n>0}^\Lambda\textrm{Tr}\left[(\hat{\tau}_0+\hat{\tau}_3)(\hat{\rho}_1+i\hat{\rho}_2)(\hat{\sigma}_0+\hat{\sigma}_3)\hat{\cal G}\right].
\end{split}
\en
Here $g_{\textrm{m}}$, $g_{\textrm{s}}$ are the interaction constants and trace over the matrix products also includes the integration over all directions of the Fermi velocity ${\mathbf v}_F=v_F\bn$. As usual, the UV-cutoff $\Lambda$ defines bare SC/SDW transition temperatures $(T_{c0},T_{s0})\sim \Lambda e^{-2/(g_{\text{s,m}}\nu_F)}$. 

In a spatially homogeneous case (\ref{Eq2Eilenr}) has a solution which does not depend on coordinates. One finds that  there exists the region 
in the values of $\Gamma_0$ where SC co-exists with the SDW state. We are interested in finding out what happens to that region in the spatially inhomogeneous case. To find a solution in a general case we use the phenomenological method proposed by Larkin \cite{Larkin1969}: we assume that the coupling constants are functions of coordinate and write them as 
\beg\label{grs}
\frac{1}{\nu_Fg_i(\br)}=\left\langle\frac{1}{\nu_Fg_i}\right\rangle+\lambda_{i}(\br), 
\en
($i=\textrm{m,s}$). The averaging is performed over disorder distributions which we assume to be Gaussian and we also assume that $\lambda_{i}\ll 1$. The inhomogeneities in the coupling constants can be characterized by the following correlation function
\beg\label{g1Corr}
\varphi_{ij}(\br-\br')=\langle \lambda_{i}(\br)\lambda_{j}(\br')\rangle, 
~\varphi_\bk=\int \varphi_{ij}(\br)e^{-i\bk\br}d^2\br.
\en
For simplicity we assume that the disorder correlators for the spin-density-wave and pairing couplings are the same. 

Our plan now consists in finding the solution of (\ref{Eq2Eilenr}) by perturbation theory.  Since $\lambda_{i}$'s are small, we seek for the  correction to the quasiclassical function due to inhomogeneities in the form:
\beg\label{SmallCorr}
\hat{\cal G}(\omega_n,\bn,\br)=\langle\hat{\cal G}(\omega_n)\rangle+\delta\hat{\cal G}(\omega_n,\bn,\br).
\en
This form implies that for the order parameters we also write $\Delta(\br)=\langle\Delta\rangle+\delta\Delta(\br)$ and 
$M(\br)=\langle M\rangle +\delta M(\br)$.
The first term on the right-hand-side (\ref{SmallCorr}) is determined by the solution of the Eilenberger equation averaged over various disorder configurations, i.e. in the spatially homogeneous case, and is given by \cite{VC-PRB2011,Khodas2020}
\beg\label{avG}
\langle\hat{\cal G}\rangle=g_{\omega}{\hat{\tau}_3\hat{\rho}_3\hat{\sigma}_0}
-f_{\omega}{\hat{\tau}_0\hat{\rho}_2\hat{\sigma}_0}+s_{\omega}{\hat{\tau}_2\hat{\rho}_3\hat{\sigma}_3}.
\en
Given the normalization condition for the function $\hat{\cal G}$, up to the linear order in $\hat{\cal G}_1$  it follows
\beg\label{G1mat}
\langle\hat{\cal G}\rangle\delta\hat{\cal G}+\delta\hat{\cal G}\langle\hat{\cal G}\rangle=0.
\en
This expression imposes a constraint on the matrix form for the function $\delta\hat{\cal G}$ and we choose to write it as follows
\beg\label{G1}
\begin{split}
\delta\hat{\cal G}=& i\alpha_x{\hat{\tau}_2\hat{\rho}_1\hat{\sigma}_3}-\beta_x{\hat{\tau}_3\hat{\rho}_1\hat{\sigma}_0}-\varsigma_x{\hat{\tau}_1\hat{\rho}_0\hat{\sigma}_3} \\&
+a_x{\hat{\tau}_3\hat{\rho}_3\hat{\sigma}_0}
+ib_x{\hat{\tau}_0\hat{\rho}_2\hat{\sigma}_0}+i\gamma_x{\hat{\tau}_2\hat{\rho}_3\hat{\sigma}_3}
\end{split}
\en
with the notation $x=(\omega_n,\bn,\br)$.
Given Eqs. (\ref{avG},\ref{G1mat}) the functions in (\ref{G1}) must satisfy $g_{\omega} a_x-if_{\omega} b_x+is_{\omega}\gamma_x=0$ and $g_{\omega}\alpha_x-is_{\omega}\beta_x-if_{\omega}\varsigma_x=0$.

The first step towards obtaining our main result is to insert  expressions (\ref{SmallCorr},\ref{avG},\ref{G1}) into (\ref{Eq2Eilenr}) and average both parts of the equation over disorder distribution function keeping the leading nonvanishing terms which contain nontrivial corrections. 
There will be three resulting equations with one of them being redundant due to the normalization condition. The remaining two equations
can be written compactly using the components of the vector ${\vec \Pi}$:
\beg\label{MFEqs0}
\Pi_zf_{\omega}-\Pi_xg_{\omega}=\langle a_x\delta\Delta \rangle, \quad \Pi_zs_{\omega}-\Pi_yg_{\omega}=\langle
a_x\delta M\rangle,
\en
where $\Pi_x=\langle\Delta\rangle+\Gamma_mf_{\omega}$, $\Pi_y=\langle M\rangle-\Gamma_ts_{\omega}$, $\Pi_z=\omega_n+\Gamma_tg_\omega$ and $\Gamma_{t,m}=\Gamma_0\pm\Gamma_\pi$. The fact that only inter-band scattering rate $\Gamma_\pi$ enters into the first equation is a manifestation of the Anderson theorem, i.e. if we set $\Gamma_\pi\to 0$, than we recover the corresponding equation for the BCS model \cite{LO-model}.

In order to compute the local (disorder-induced) correlation functions featured in Eqs. (\ref{MFEqs0}), we go back to the Eilenberger equation (\ref{Eq2Eilenr}) and keep the terms up to the first order in the components of $\delta\hat{\cal G}$. The solution of the Eilenberger equation can be conveniently found by going into momentum representation  
\beg\label{Eqskrep}
{\begin{split}
a_k&=-\frac{\Pi_z[f_{\omega}\delta\Delta(\bk)+s_{\omega}\delta M(\bk)]}{(v_F/2)^2(\bn\bk)^2+{\vec \Pi}^2}, \\
ib_k&=\frac{\Pi_yf_{\omega}\delta M(\bk)-\delta\Delta(\bk)(\Pi_zg_{\omega}+\Pi_ys_{\omega})}{(v_F/2)^2(\bn\bk)^2+{\vec \Pi}^2}, \\
i\gamma_k&=-\frac{\Pi_yf_{\omega}\delta\Delta(\bk)-\delta M(\bk)(\Pi_zg_{\omega}+\Pi_xf_{\omega})}{(v_F/2)^2(\bn\bk)^2+{\vec \Pi}^2},
\end{split}}
\en
where now $k=(\omega_n;\bn,\bk)$
It is easy to check that these relations satisfy corresponding constraint condition. In order to find the expressions which are valid for an arbitrary values of $k_Fl$ ($k_F$ is a Fermi momentum and $l$ is the mean-free path), in (\ref{Eqskrep}) one needs to replace 
$\delta\Delta(\bk)\to\delta\Delta(\bk)-i\Gamma_m\langle b_k\rangle_\bn-({f_{\omega}}/{g_{\omega}})\Gamma_t\langle a_k\rangle_\bn$, 
$\delta M(\bk)\to \delta M(\bk)-i\Gamma_t\langle \gamma_k\rangle_\bn-({s_{\omega}}/{g_{\omega}})\Gamma_t\langle a_k\rangle_\bn$
(here $\langle...\rangle_\bn$ denotes the averaging over all directions of $\bn$) and solve (\ref{Eqskrep}) for $\langle a_k\rangle_\bn$, $\langle b_k\rangle_\bn$
and $\langle \gamma_k\rangle_\bn$ after averaging them over $\bn$. 
All these expressions can be found in the closed form. 
Lastly, we note that the expressions for the remaining three functions $\alpha_k$, $\beta_k$ and $\varsigma_k$ are of no importance to us since they do not contribute to the self-consistency equations for their averages over the directions of the Fermi velocity vanish identically. 

With the help of the first equation (\ref{Eqskrep}) we can now express the disorder correlation functions (\ref{MFEqs0}) in terms of the order parameter correlators. For brevity, we represent it in terms of the two-component field $\hat{\Phi}(\br)=[\delta\Delta(\br),\delta M(\br)]$: 
\beg\label{OPCorrs}
\langle\hat{\Phi}(\br)\hat{\Phi}(\br')\rangle=\int\frac{d^2\bk}{(2\pi)^2}\left[
\begin{matrix} {\cal D}_\bk &  {\cal C}_\bk \\ 
{\cal C}_\bk & {\cal M}_\bk
\end{matrix}\right]e^{i\bk(\br-\br')}.
\en
In their turn, the correlators (\ref{OPCorrs}) can be expressed in terms of the correlators of the interactions constants (\ref{g1Corr}) by solving the following system of linear equations derived from the self-consistency conditions: 
\beg\label{SelfAgain}
\begin{split}
&\pi T\sum\limits_{\omega_n>0}^\infty\left(\frac{g_\omega}{\omega_n+2\Gamma_tg_\omega}+\chi_ys_\omega-{\vec p}_\omega{\vec \chi}\right)\delta M(\bk)\\&+
\pi T\sum\limits_{\omega_n>0}^\infty\chi_yf_\omega\delta\Delta(\bk)=-\langle M\rangle \lambda_{\textrm{m}}(\bk), \\
&\pi T\sum\limits_{\omega_n>0}^\infty\left(\frac{g_\omega}{\omega_n+2\Gamma_\pi g_\omega}+\chi_xf_\omega-{\vec p}_\omega{\vec \chi}\right)\delta\Delta(\bk)\\&+
\pi T\sum\limits_{\omega_n>0}^\infty\chi_yf_\omega \delta M(\bk)=-\langle\Delta\rangle \lambda_{\textrm{s}}(\bk).
\end{split}
\en
Here functions $\chi_\alpha$ are the components of the vector ${\vec \chi}=(\chi_x,\chi_y,\chi_z)$ with ${\chi}_j=(\Pi_j/|{\vec \Pi}|)[(v_Fk/2)^2+{\vec \Pi}^2]^{-1/2}$ and ${\vec p}_\omega=(f_{\omega},s_{\omega},g_{\omega})$. 

In what follows, we are primarily interested in finding how inhomogeneity-induced correlations influence the co-existence region. For this purpose, we only need to analyze the critical temperatures $T_c(M)$, which determines the onset of the SC emerging from the preexisting SDW state, and $T_s(\Delta)$ which sets the boundary between the coexistence region and purely SC state in the temperature-doping phase diagram. Therefore, we only need to analyze the expressions for the disorder correlators when one of the order parameters is zero. 

\paragraph{Results for $T_c(M)$ boundary.} In this case $\langle M\rangle\not=0$,~$\langle\Delta\rangle=0$, and we also set $\delta \Delta=0$, which means that the first equation (\ref{MFEqs0}) is satisfied identically, while the second equation can be parametrized as follows
\beg\label{Eq4TcM}
\Pi_zs_{\omega}-\Pi_yg_{\omega}=-\eta_\textrm{m}\langle M\rangle g_{\omega}s_{\omega},
\en
where we parametrized the correlator as $\left\langle a_x \delta M\right\rangle=-\eta_\textrm{m}\langle M\rangle g_{\omega}s_{\omega}$.
The expression for the parameter $\eta_\textrm{m}$ which is applicable for arbitrary values of $k_Fl$ is:
\beg\label{ETAM1}
\eta_\textrm{m}=\langle M\rangle\int\limits_0^{\infty}
\frac{d^2\bm{k}}{(2\pi)^2}\frac{{\vec p}_\omega{\vec \chi}{\cal M}_{\bk}}{(1+2\Gamma_t\chi_z-\Gamma_t{\vec p}_\omega{\vec \chi})},
\en
where we used the identity ${\vec p}_\omega{\vec \chi}=\chi_z/g_\omega$, introduced  
\beg\label{GiveMk}
\begin{split}
{\cal M}_\bk=\varphi_\bk\left[\pi T_c\sum\limits_{\omega_n>0}^\infty\left(\frac{g_\omega}{\omega_n+2\Gamma_tg_\omega}+\chi_ys_\omega-{\vec p}_\omega{\vec \chi}\right)\right]^{-2}
\end{split}
\en
and rescaled ${\cal M}_\bk\to\langle M\rangle^2{\cal M}_\bk$. Without loss of generality, we consider
\beg\label{phir}
\varphi_\bk={\varphi_0}e^{-(kr_c/2)^2},
\en
where $\varphi_0$ and $r_c$ are the phenomenological parameters characterizing the magnitude and scale of the inhomogeneities. Although both $\langle M\rangle$ and $\eta_m$ must be solved for self-consistently, from Eq. (\ref{Eq4TcM}) we see that inhomogeneities produce the shift in the scattering rate $2\Gamma_t\to 2\Gamma_t+\eta_\textrm{m}\langle M\rangle$. This means that at least at very small values of $\eta_{\text{m}}$, $T_c(M)$ must increase compared to its value in the spatially homogeneous case for suppression of $\langle M\rangle$. Qualitatively this implies boost for superconductivity. The actual magnitude of the parameter $\eta_{\textrm{m}}$ crucially depends on the correlation radius $r_c$: when the correlation radius $k_Fr_c\sim 1$ and $r_c\ll v_F/\langle\Delta\rangle$ we expect $\eta_{\textrm{m}}\ll 1$. 

\paragraph{Results for $T_s(\Delta)$ boundary.} In this case $\langle \Delta\rangle\not=0$,~$\langle M\rangle=0$, and thus we have
\beg\label{Eq4TsD}
\Pi_zf_{\omega}-\Pi_xg_{\omega}=-\eta_\textrm{s}\langle \Delta\rangle g_{\omega}f_{\omega},
\en
where the dimensionless parameter $\eta_{\textrm{s}}$ is given by
\beg\label{ETAS1}
\eta_\textrm{s}\approx\langle\Delta\rangle\int\limits_0^{\infty}
\frac{d^2\bm{k}}{(2\pi)^2}\frac{{\vec p}_\omega{\vec \chi}{\cal D}_{\bk}}{(1-\Gamma{\vec p}_\omega{\vec \chi})}
\en
and the rescaled ${\cal D}_\bk\to \langle\Delta\rangle^2{\cal D}_\bk$ correlator is 
\beg\label{GiveDk}
\begin{split}
{\cal D}_\bk=\varphi_\bk\left[\pi T_s\sum\limits_{\omega_n>0}^\infty\left(\frac{g_\omega}{\omega_n+2\Gamma_\pi g_\omega}+\chi_xf_\omega-{\vec p}_\omega{\vec \chi}\right)\right]^{-2}.
\end{split}
\en
We note that expression (\ref{ETAS1}) acquires such a simple form only when we assume that $\Gamma_\pi\ll \Gamma_0=\Gamma$. This approximation is not restrictive as $\Gamma_\pi$ is primarily responsible for bending SC dome of $T_c(\Gamma_\pi)$ at larger dopings, and has weaker influence of the physics near the coexistence region. In addition, we observe again that inhomogeneities lead to an increase in the inter-band scattering rate $2\Gamma_\pi\to 2\Gamma_\pi+\eta_\textrm{s}\langle\Delta\rangle$, so it is not a priory clear whether it will yield the suppression or boost of $T_s(\Delta)$. To resolve this question we need to employ a self-consistent approach. 

\begin{figure}[t!]
  \centering
 \includegraphics[width=3.25in]{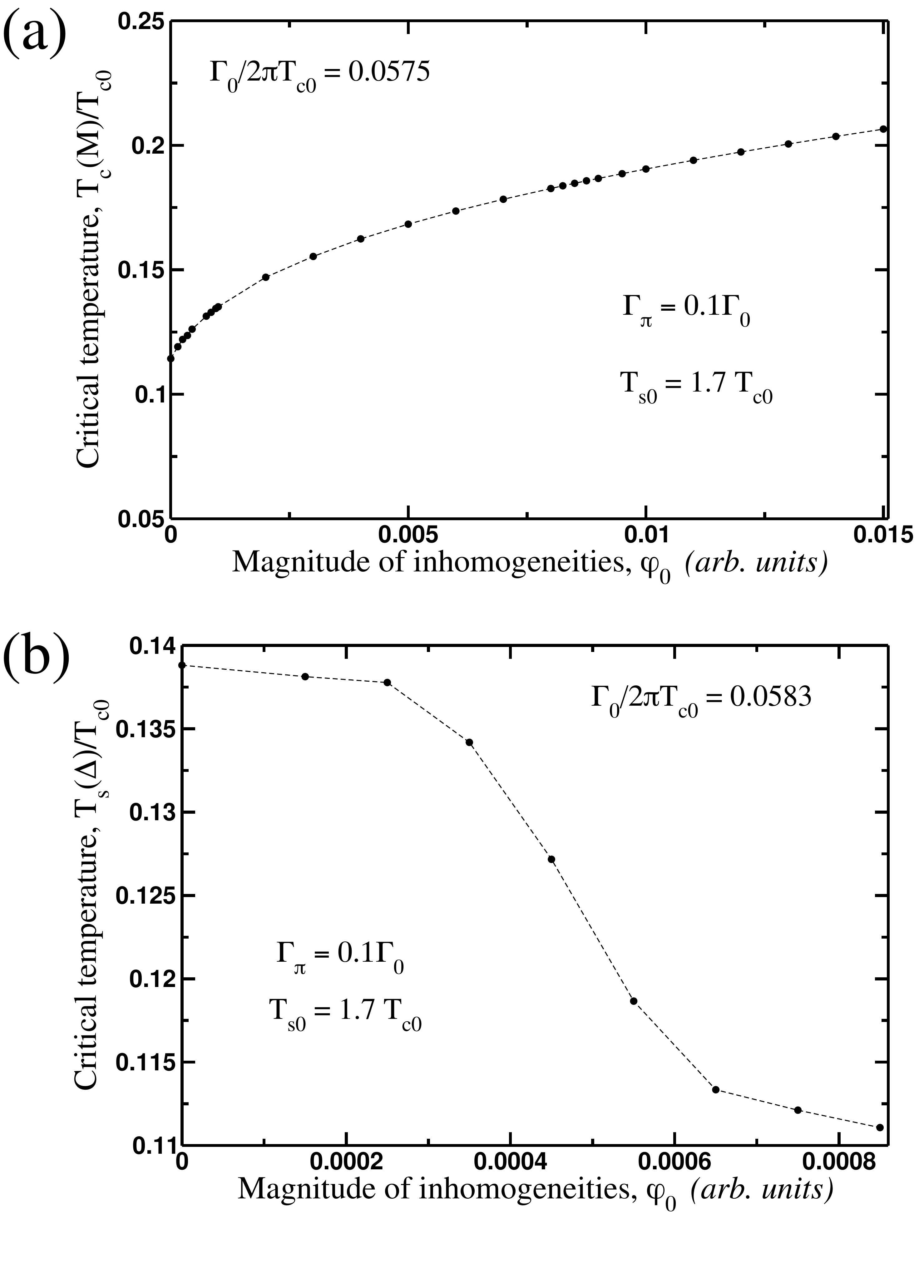}
  \caption{Panel (a): results of the self-consistent solution of the equations (\ref{Eq4TcM},\ref{ETAM1}) for the superconducting critical temperature inside the SDW phase. Panel (b): results of the self-consistent solution of the equations (\ref{Eq4TsD},\ref{ETAS1}) for the SDW transition temperature inside the superconducting phase. These results have been obtained by neglecting the dependence of the parameters $\eta_{\textrm{m,s}}$ on Matsubara frequency.}\label{Fig2}
\end{figure}

\paragraph{Self-consistent method.} Parameters $\eta_{\textrm{m,s}}$ are functions of Matsubara frequency and, therefore, equations (\ref{ETAM1},\ref{ETAS1}), must be solved self-consistently with (\ref{Eq4TcM},\ref{Eq4TsD}) along with the equation for $\langle\Delta\rangle$ and $\langle M\rangle$. However, in the case of strong inhomogeneities the main contribution to the integral comes from the region of momenta $k\sim r_c^{-1}$ ($r_c$ is the disorder correlation radius) and the frequency dependence of these parameters can be neglected. In Fig. \ref{Fig2} we show the results of the self-consistent solution of the equations above for the critical temperature $T_c(M)$ and $T_s(\Delta)$ correspondingly as functions of parameter $\varphi_0$. As we have expected, $T_c(M)$ increases with increase with the magnitude of inhomogeneities, while $T_s(\Delta)$ decreases with the increase in $\varphi_0$. This means that within the linear approximation we have adopted, spatial inhomogeneities have a much more profound effect on the magnetic transition than on superconductivity. 

\paragraph{Summary and discussion.} 

In conclusion, we have considered the impact of spatial pairing-potential correlations induced by short-scale disorder fluctuations on the interplay of the SDW-SC competition in multiband metals. We found that quantitative effects stemming from the physics of short scales are the enhancement of superconducting $T_c$ in the optimally doped region and widening of the coexistence phase. These conclusions are rather robust and fairly  universal as microscopic form of the disorder correlation function is not essential. It is only the correlation radius and strength of correlations that determine relevant parameters of the model.     

The extent of results presented in this study is limited by two major factors. First, we considered only a minimal two-band model. A more elaborate treatment will bring additional features most notably possible disorder-induced topological change of the superconducting gap structure \cite{MizukamiNC2014,ChoSciAdv2016}, appearance of a narrow dome of $s+is'$ time-reversal broken superconductivity separating gapped and nodal regions \cite{GrinenkoNP2020}, as well as effects of nematic correlations \cite{ZhengPRL2018}. All these phenomena have profound observed experimental signatures. However, these complications do not change the main conclusion of this work concerning the effect of short-range disorder fluctuations on the width of coexistence region. Indeed, multi-band character simply brings additional renormalizations of $\Gamma_\pi$, thus steeper suppression of $T_c$ in the overdoped region, but has qualitatively the same weaker effect in the domain of optimal doping, as supported by our numerical self-consistent analysis. These details can be further tackled quantitatively based on the quasiclassical theory of three-band modeling of magnetic order in iron-pnictides \cite{MDMK} extended to superconducting scenarios. Second, we considered only weak impurities treated at the level of Born approximation, thus missed physics of induced Yu-Shiba-Rusinov localized bound or miniband subgap states \cite{Yu,Shiba,Rusinov} that can be captured by a full $\hat{T}$-matrix analysis. This is still an open problem to address in the context of SDW-SC coexistence and density of states subgap structure that we leave for further investigation.      

\paragraph*{Acknowledgments.}
This work was financially supported by the National Science Foundation grant NSF-DMR-2002795 (M.D.) and by the U.S. Department of Energy (DOE), Office of Science, Basic Energy Sciences (BES) Program for Materials and Chemistry Research in Quantum Information Science under Award No. DE-SC0020313 (A.L.). 

\bibliography{main1biblio}

\begin{thebibliography}{10}
\bibitem{MinusSign}
The minus sign in front of $-\hat{c}_{-\bp\dn}$ and $-\hat{f}_{-\bp\dn}$ is
  important for the correct definition of the spin operator ${\vec S}(\br)$ to
  ensure its proper transformation under the time-reversal symmetry operator.
\end{thebibliography}

\end{document}